\documentclass[aps,12pt,tightenlines]{revtex4}
\usepackage[dvips]{color,graphicx}
\begin{document}
\newcommand{\boldsigma}{\mbox{\boldmath $\sigma$}}
\newcommand {\boldgamma}{\mbox{\boldmath$\gamma$}}
\newcommand{\boldtau}{\mbox{\boldmath $\tau$}}
\newcommand{\bftau}{\mbox{\boldmath $\tau$}}
\newcommand{\bfx}{{\bf x}}
\newcommand{\bfy}{{\bf y}}
\newcommand{\bfP}{{\bf P}}
\def\poinc{Poincar\'{e} }
\newcommand{\eq}[1]{Eq.~(\ref{#1})}
\def\bfq {{\bf q}}
\def\bfm {{\bf m}}
\def\bfs {{\bf s}}
\def\bfn {{\bf n}}
\def\bfqp {{\bf q}_\perp}
\def\bfK{{\bf K}}
\def\bfKp{{\bf K}_\perp}
\def\bfL{{\bf L}}
\def\bfk{{\bf k}}
\def\bfp{{\bf p}}  
\newcommand{\bfkap}{\mbox{\boldmath $\kappa$}} 
\def\bfr{{\bf r}} 
\def\bfy{{\bf y}} 
\def\bfx{{\bf x}} 
\def\be{\begin{equation}}
 \def \ee{\end{equation}}
\def\bea{\begin{eqnarray}}
  \def\eea{\end{eqnarray}}
\newcommand{\eqn} {Eq.~(\ref )}
% braket related commands
% bra-ket = b-k = <>
\newcommand{\bb}{\langle}
\newcommand{\kk}{\rangle}
\newcommand{\bk}[4]{\bb #1\,#2 \!\mid\! #3\,#4 \kk}
\newcommand{\kb}[4]{\mid\!#1\,#2 \!\mid}

\def\notp{{\not\! p}}
\def\notk{{\not\! k}}
\def\up{{\uparrow}}
\def\down{{\downarrow}}
\def\bfb{{\bf b}}

%\documentstyle[aps]{revtex4}
%\tighten
\setlength{\textheight}{8.60in}
\setlength{\textwidth}{6.6in}
\setlength{\topmargin}{-.40in}
\setlength{\oddsidemargin}{-.125in}
\tolerance=1000
\baselineskip=14pt plus 1pt minus 1pt

\def\poinc{Poincar\'{e} }
\def\bfq {{\bf q}}
\def\bfK{{\bf K}}
\def\bfL{{\bf L}}
\def\bfk{{\bf k}}
\def\bfp{{\bf p}}  
\def\be{\begin{equation}}
 \def \ee{\end{equation}}
\def\bea{\begin{eqnarray}}
  \def\eea{\end{eqnarray}}
\def\eqn {Eq.~(\ref )}
% braket related commands
% bra-ket = b-k = <>

\newcommand{\kx}[2]{\mid\! #1\,#2 \kk}
\def\notp{{\not\! p}}
\def\notk{{\not\! k}}
\def\up{{\uparrow}}
\def\down{{\downarrow}}
\def\bfb{{\bf b}}

\vspace{1.0cm}
\vspace{.50cm}
\title{\begin{flushright}{\normalsize NT@UW-06-07}\end{flushright}
Shapes of the Nucleon}

\author{Alexander  Kvinikhidze}
%$a$
\affiliation{The Mathematical Institute of Georgian Academy of Sciences, Tbilisi, Georgia} 
\author{Gerald A. Miller}
\affiliation{ University of Washington
  Seattle, WA 98195-1560}

\sloppy

%\begin{document}

%\maketitle
\begin{abstract} Previously defined spin-dependent quark densities that are matrix elements
of specific 
 density operators in 
proton states of definite spin-polarization generally  have 
an infinite 
variety of non-spherical shapes. The present application is concerned with
both charge and matter densities. We show that the  Gross \& Agbakpe model
nucleon 
harbors an interesting variety of 
non-spherical shapes.
\end{abstract}
\maketitle
\vskip0.5cm

\section{Introduction}
Recent data \cite {Jones:1999rz,Gayou:2001qd} 
showing that  the ratio of the proton's electric and magnetic form
factor $G_E/G_M$, falls with increasing momentum transfer $Q^2$ for 
1$<Q^2<6 $ GeV$^2$ and equivalently that 
$QF_2(Q^2)/F_1(Q^2)$ is a approximately constant
have created considerable attention. This behavior  
indicates that the sum of the orbital angular momentum of the 
quarks in the proton is non-vanishing
\cite{ralston,Braun:2001tj,Miller:2002qb,Ji}.

It is natural to consider the connection between orbital angular momentum
$(l)$ and the 
 shape of the nucleon, and one of us showed  \cite{Miller:2003sa},   using 
the  proton  model of
Ref.~\cite{Frank:1995pv},
  that the   rest-frame ground-state 
matrix elements of spin-dependent
density operators
reveal a host of non-spherical shapes. 
The use of the spin-dependent density operator is the key feature
that allows the detailed connection between orbital, spin and total
angular momentum to be revealed in quantum systems.
Matrix elements of the non-relativistic spin-density operator 
have been measured in condensed
matter systems\cite{Prokes}, revealing the orbital angular momentum content  of
electron orbitals.

In the model of \cite{Frank:1995pv}  the relativistic nature of the
quarks enters via their lower components of  Dirac spinors that are part of the
wave function. 
It is natural to ask if  is it necessary that  the presence of 
 relativistically
moving quarks always causes a non-spherical shape (as determined by matrix elements
of spin-dependent density operators).
We shall not address this general question  here, and  consider only the 
model dependence of the shape. In particular, 
Gross \& Agbakpe (GA)\cite{ga} 
have claimed  to construct a manifestly covariant
nucleon wave function that contains only $l=0$ such that 
``{\it the nucleon would still be spherical''} and also describe the measured nucleon electromagnetic
form factors. We show that this contention is not correct by
evaluating
the matrix elements of the spin-dependent density 
operator %$\widehat{\rho}_{\cal O}(\bfK,\bfn)$ 
using the
GA wave function.

\section{Spin-Dependent Density Operators}
We begin by explaining how  
the  shapes of a nucleon are  exhibited 
by studying the rest-frame ground-state 
matrix elements of spin-dependent
density operators.  
The usual charge density operator in non-relativistic
quantum mechanics is given by 
\bea \widehat{\rho}(\bfr)= \sum_i {e_i\over e}
\delta(\bfr-\bfr_i)\eea
where $e_i/e$ is the charge of the $i$'th particle 
(in units of the proton charge) and $\bfr_i$ its position
operator. Matrix elements of this operator yield the charge density of
a system. Suppose the particles also have spin 1/2. Then one can measure
the probability that particle is at a given position $\bfr$ and has a 
spin in an arbitrary, fixed direction specified by a unit vector
 $\bfn.$ The  spin projection operator
is $(1+\boldsigma\cdot\bfn)/2 $, so the  spin-dependent density operator
is 
\bea \widehat{\rho}(\bfr,\bfn)= \sum_i {e_i\over e}
\delta(\bfr-\bfr_i){1\over2}(1+\boldsigma\cdot\bfn).\label{sddr}\eea 
The 
 spin-dependent density allows 
the presence of the orbital angular momentum to be revealed in
 the shape of the computed density. 
Matrix elements of the spin-density operator of \eq{sddr}
have been measured in condensed
matter systems\cite{Prokes}, revealing the orbital angular momentum content  of
electron orbitals.

It is worthwhile to consider a simple example of a
 single charged particle
moving in a fixed rotationally invariant potential in a state of quantum
numbers: $n,1,1/2,m$ we find
\bea \rho(\bfr,\bfn)=\langle n,1,1/2,m\left\vert \widehat{\rho}(\bfr,\bfn)\right\vert n,1,1/2,m
\rangle
={R_{n,1,1/2}^2(r)\over 2}\bb  
m\vert1+2\boldsigma\cdot \hat{\bfr}\;{\bfn} \cdot \hat{\bfr} -\boldsigma\cdot
{\bfn}\vert m\kk.\eea
Suppose  $\hat{\bfn}$
 is either parallel or anti-parallel to
the direction of the proton angular momentum defined by the vector 
$\hat\bfs$. The direction of the
 vector $\hat\bfs$ defines an axis (the ``z-axis''), and
 the direction of vectors can be represented in terms of this axis:
$\hat{\bfs}\cdot\hat{\bfr}=\cos\theta$. With this notation
$ \rho(\bfr,{\bfn}=\hat{\bfs})={R_{n,1,1/2}^2(r)}\cos^2\theta,\;
 \rho(\bfr,{\bfn}=-\hat{\bfs}  )={R_{n,1,1/2}^2(r)}\sin^2\theta$ and
 the non-spherical shape is exhibited. 
The average of these
two cases   is  
a spherical shape as is the  average over the direction of $\hat\bfs$.
The spherical shape claimed in \cite{ga} arises from taking an average
over the direction of $\bfn$. Averaging over this  direction necessarily
buries the effects of orbital angular momentum responsible for non-spherical
shapes.

The  density we have discussed so far is defined
in terms of position, but when quantum field theory applies it is convenient to 
define similar operators that give
the probability for a particle to have a given momentum, $\bfK$, 
 and a given direction
of spin, $\bfn$.
The field-theoretic version of the spin-dependent charge 
 density operator is \cite{Miller:2003sa}
\bea\widehat{\rho}_{\cal O}(\bfK,\bfn)=\int {d^3r\over(2\pi)^3} e^{i\bfK\cdot\bfr}%\langle N\vert
\bar{\psi}(\bfr){\cal O}
(\gamma^0+\boldgamma\cdot\bfn\gamma_5)\psi({\bf 0}), %\vert N\rangle
\label{qft}\eea 
where %$\vert N\rangle$ represents the nucleon, and 
 $\cal O$ is  
$\widehat{Q}/e$, the quark charge operator in units of the proton charge.
Here we shall also consider the case in which ${\cal O}=1$. 
Its matrix element
along with \eq{qft} gives the spin-dependent matter densities.
The quark  field
operators are evaluated at equal time. There are no gluon fields in the 
relativistic constituent quark models used here. The question of color gauge
invariance of QCD will be taken up elsewhere. The matrix 
element of this density operator in
a nucleon state of definite total angular momentum  defined
 by the unit vector $\bfs,\;\vert\Psi_{\bfs}\kk$  is
\bea
\rho_{\cal O}(\bfK,\bfn,\bfs)
\equiv\bb\Psi_\bfs\vert \widehat{\rho}_{\cal O}(\bfK,\bfn)\vert\Psi_{\bfs}\kk
,\label{sdd}\eea
where the subscript ${\cal O} =Q,\;1$ specifies the operator used in \eq{qft}.

The most general shape of the proton, 
obtained if parity and rotational invariance are upheld is
\bea
\rho_{\cal O}(\bfK,\bfn,\bfs)=A_{\cal O}(\bfK^2)+B_{\cal O}(\bfK^2)\bfn\cdot\bfs+C_{\cal O}
(\bfK^2)\bfn\cdot\bfK\;\bfs\cdot\bfK \label{genshape}
,\eea
with the last term generating the non-spherical shape. Any 
wave function that yields a non-zero value of the coefficient $C_{\cal O}(\bfK^2)$ 
represents a system
of a non-spherical shape.

In Ref.~\cite{Miller:2003sa} the spin-dependent charge density of 
a relativistic three-quark constituent model \cite{Frank:1995pv}
was evaluated with the result (for the proton):
\bea
\rho_Q^p(\bfK,\bfn,\bfs) \;%\bb\Psi_s\vert \hat{\rho}(\bfK,\hat{\bfn})\vert\Psi_{s}\kk
=\rho(K){1\over2}(1+\bfn\cdot\hat{\bfs}+{\gamma(K)}
(1-\bfn\cdot\hat{\bfs}+2\hat{\bfK}\cdot\bfn\hat{\bfK}\cdot\hat{\bfs}))
\label{shapek} \eea 
with \bea \rho(K)\equiv\int\;d^3k \Phi^2(k,K)(E(K)+m),\;\gamma(K)\equiv
{E(K)-m\over E(K)+m},\label{rhodef}\eea
with $\Phi(k,K)$ the  three-quark wave function and $m$ the constituent quark mass
specified in 
 \cite{Miller:2003sa}. The result for the neutron is 
\bea
\rho_Q^n(\bfK,\bfn,\bfs) \;%\bb\Psi_s\vert \hat{\rho}(\bfK,\hat{\bfn})\vert\Psi_{s}\kk
=\rho(K){1\over18}(1-\bfn\cdot\hat{\bfs}+{\gamma(K)}
(1+\bfn\cdot\hat{\bfs}-2\hat{\bfK}\cdot\bfn\hat{\bfK}\cdot\hat{\bfs}))
\label{shapekn} \eea 

The shape for a given value of $K$ is determined by the ratio
$\gamma(K)$ which reaches \cite{Miller:2003sa} a value of 0.6 for $K=1$ GeV/c.
This implies considerable non-sphericity. 
The probability that a given value of $K$
 is determined by the function
$K^2\rho(K)$, displayed in Fig.~1 of
\cite{Miller:2003sa}. The most likely value of $K\approx 0.25$ GeV/c corresponding 
to $\gamma(K)=0.16$.

Some special cases of  Eq.~(\ref{shapek}) are interesting.
Suppose the  quark spin is parallel to the proton spin,  $\bfn =
\hat{\bfs}$,
then
$\hat{\rho}_Q(K,\bfn =\hat{\bfs})=\rho(K)
\left(1+\gamma(K) \cos^2\theta\right)$. 
For small $K$ the shape is nearly spherical,
 but for large $K$ the $\cos^2\theta$ term becomes prominent. On the other
hand, the quark spin could be anti-parallel to the proton spin, 
 $\bfn =-\hat{\bfs}$. Then we find:
$\hat{\rho}(K,\bfn =-\hat{\bfs})=\rho(K)\gamma(K)
 \sin^2\theta,$ 
and the shape is that of a torus.
%for all values of $K$. 
We may also take the
 quark spin perpendicular to the proton spin 
$\bfn\cdot\bfs=0$, so that
$
\rho_Q(\bfK,\bfn\cdot\bfs=0)=\rho  (K) (1+\gamma(K))/2 +\gamma(K)
\sin\theta\cos\theta (\cos\phi n_x+\sin\phi n_y),$
to display the dependence on the azimuthal angle.
In each case, the non-spherical nature arises from the term proportional 
to $\gamma(K)$ caused by the lower components of the 
Dirac spinor that is part of the proton wave function.
For the neutron one obtains a doughnut 
shape for $\bfn =\hat{\bfs}$, and a peanut for
$\bfn =-\hat{\bfs}$.  Plots of these shapes have appeared in
Ref.~\cite{Miller:2003sa} as well as in the NY Times\cite{nyt}.

For the matter distribution $({\cal O}=1)$ we find
\bea
\rho_1(\bfK,\bfn,\bfs) %\bb\Psi_s\vert \hat{\rho}(\bfK,\hat{\bfn})\vert\Psi_{s}\kk
=\rho(K){1\over2}(3+\bfn\cdot\hat{\bfs}+{\gamma(K)}
(3-\bfn\cdot\hat{\bfs}+2\hat{\bfK}\cdot\bfn\hat{\bfK}\cdot\hat{\bfs})).
\label{mshapek} \eea 
For a fixed value of $K$, 
this density looks more like a sphere than that of \eq{shapek}.
For example, if $\bfn=-\bfs$ one finds 
$\rho_1(\bfK,\bfn)=\rho(K)\left(1+\gamma \sin^2\theta\right)$, 
which is mainly spherical instead of toroid.
Similarly if $\bfn=\bfs$ one finds 
$\rho_1(\bfK,\bfn)=\rho(K)\left(2+\gamma \cos^2\theta\right)$.
For the case that $\bfn\cdot\hat{\bfs}=0$, with $\bfn$ in the $y$-direction
we find 
$\rho_1(\bfK,\bfn)=\rho(K)\left[3/2+\gamma(3/2+ \cos\theta\sin\theta\cos\phi)\right]$.
%\vskip3.0cm
\begin{figure} [t!]
%\begin{Large}
\unitlength1.1cm
\begin{picture}(10,8)(0,1.1)
\includegraphics{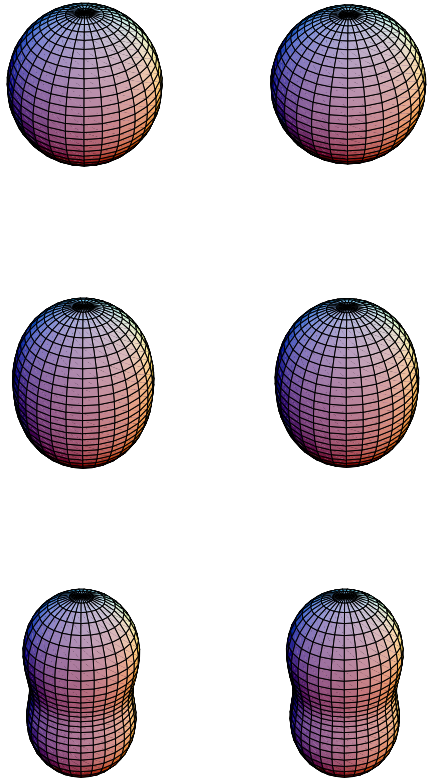}
\end{picture}
%\end{Large}
\label{fig:para}
\caption{Matter distributions. The spin direction $\bfs$ is taken as  vertical,
$\bf{\hat {z}}$, with
$\bfn=\hat{\bfs}$. Right column model of \cite{Miller:2003sa}  
, left column
model of \cite{ga}. First row $K$ = 250 MeV/c, second row $K$=1 GeV/c, 
third row $K\to \infty$.}
\end{figure}

\section{Model of Gross \& Agbakpe}

In Ref.~\cite{ga}
 explicit effects of quark spinors are avoided 
by removing a 
factor of the single-quark free propagator that would appear in a solution of the
Bethe-Salpeter equation.
The GA nucleon wave function consists of two terms which involve either 
a scalar or vector diquark. The  diquark covariant polarization vector is $\eta^\mu$.
The relativistic impulse approximation is used to evaluate the matrix element of the 
electromagnetic current operator. In the evaluation of \cite{ga} the 
elecromagnetic matrix element, 
${\cal J}^\mu_I$ for a nucleon of isospin $I$ and four-momentum transfer $q$ is
three times the integrated
current of a single off-shell quark, computed 
with the spectator diquark system on mass-shell:
%\bea
%\Psi_{_\alpha} &=&\frac{1}{\sqrt{2}} u_{_{\alpha}}(P,s)\, \xi_{_{}}^{0}\,\chi^I \;\psi_0(P,p) \cr
%&+&\frac{1}{3\sqrt{2}} (\gamma_5 \gamma_\mu)_{_{\alpha\beta}}\,\eta^{\mu} u_{_{\beta}}(P,s) 
%\left(\tau\cdot\xi^{1}_{_{}}\right)\chi^I_{}\psi_1(P,p) \qquad
%\label{eq4}
%\eea
%
%
\bea
{\cal J}_{I}^\mu =\frac{3}{2}\int \frac{m_s^3\,d^3
{\kappa}}{(2\pi)^3 m_s 2E_s(\kappa)} \Bigg\{j_{I}^\mu \psi_0(P_+,p)\psi_0(P_-,p)\qquad\cr
 -\frac{1}{9} \gamma_\nu \gamma_5 \tau_j j_{I}^\mu
\tau_j \gamma_5\gamma_{\nu'} \Delta^{\nu\nu'}\!
\psi_1(P_+,p)\psi_1(P_-,p)
\Bigg\} ,\quad \label{eq8}
\eea
%
%\end{widetext}
where  $P_\pm\equiv P\pm \frac{1}{2}q$.   
The CQ current, $j_{I}^\mu$, is specified in \cite{ga}. In deriving \eq{eq8}
the polarization 
vectors of the diquark are summed over, using
$%\bea
\sum_\eta \eta^{\nu*} \eta^{\nu'}\equiv \Delta^{\nu\nu'} = 
-g^{\nu\nu'} + \frac{p^\nu p^{\nu'}}{m_s^2}\, .
$%\eea
%$\chi^{0,1}$ are mixed anti-symmetric or symmetric di-quark isospin wave functions, 
%$\chi^I$ specifies a proton $(I=1/2)$ or neutron $(I=-1/2)$,
The quantity $p$ is the on-shell diquark four-momentum ($p^2=m_s^2$), and  $u(P,s)$ 
is the spinor of the nucleon. %, $\alpha$ is the Dirac index of the 
%off-shell quark, and $\eta$ is the covariant polarization 
%vector of the spin-one diquark. 
The scalar functions $\psi_{0,1}$ describing the nucleon with scalar (0) and vector (1) diquark
systems  are chosen to  depend  only on the
 variable $\chi=((M-m_s)^2-(P-p)^2)/(Mm_s).$
For completeness we present the functions $\psi_{0,1}$:
\bea
\psi_0(P,p)=\frac{N_0}{m_s(\beta_1-2 +\chi)(\beta_2-2 +\chi)},
\quad
\eea
where $\beta_1$ and $\beta_2$ are range parameters in units of $Mm_s$, $N_0$ is a 
normalization constant and 
\bea
\frac{\psi_1(P,p)}{\psi_0(P,p)}=\sqrt{6\over 2+(\chi+2)^2}\equiv{\cal R} \;
.  \label{BA}
\eea
We are concerned only with rest-frame matrix elements. In this frame  
 $\chi=2(E_s(\kappa)-1)$, 
where $E_s(\kappa)=\sqrt{1+\kappa^2}$ where
$\kappa=\sqrt{1+\bfp^2/m_s^2}$.

GA claim  that their  wave function  depends only on the magnitude of ${\bf p}$, 
and  therefore  is spherically symmetric. This is not correct. While the functions
$\psi_{0,1}$ depend only on the magnitude of ${\bf p}$, 
the nucleon wave function %$\Psi_\alpha$
contains the polarization vector $\eta^\mu$ and this yields a non-spherical shape as
defined by 
taking the matrix elements of the operator $\widehat{\rho}_{\cal O}(\bfK,\bfn).$
Before proceeding we note that 
the relation between the di-quark momentum $\bfp$ of GA is simply the negative of 
the quark momentum $\bfK$
of \cite{Miller:2003sa}: $\bfp=-\bfK.$

The expectation value of  the operator $\widehat{\rho}_{\cal O}(\bfK,\bfn)$ in the
nucleon of total 
angular momentum $\bfs$ of the GA model (corresponding to \eq{sdd})
 is defined  as ${\rho}_{\cal O}^{GA}
(\bfK,\bfn,\bfs)$. Once again  the result is three times
the integrated current of a single off-shell quark, calculated with the 
spectator ``diquark'' system on mass-shell. A straightforward evaluation yields
\bea 
{\rho}_{\cal O}^{GA}(\bfK,\bfn,\bfs)&=&{3\over 2E(\bfK^2)}
\bar{U}[{\cal O}_0(\gamma_0+\bfn\cdot\boldgamma\gamma_5)
\vert\psi_0(\bfK)^2\vert\nonumber\\
&&+{1\over9}\tau_j{\cal O}_1\tau_j\gamma_\mu(\gamma^0+\bfn\cdot\boldgamma\gamma_5)
\gamma_\nu\left({p^\mu p^\nu\over m_s^2}-g^{\mu\nu}\right)
\vert\psi_1(\bfK)\vert^2] U,\eea
where $U$ is the Dirac spinor for a 
nucleon  of total angular momentum $\bfs$ at rest, and
$E(\bfK^2)=\sqrt{\bfK^2+m_s^2}.$
For the charge density: 
${\cal O}_0 =$ 2/3(proton),$\;$ -1/3 (neutron), $\tau_j{\cal O}_1\tau_j$=0 (proton),-1 
(neutron). For the matter density:
${\cal O}_0 =1,\;\tau_j{\cal O}_1\tau_j=3.$ 
Thus we explicitly
find 
\bea &&{\rho}_Q^{GAp}(\bfK,\bfn,\bfs)=
{1\over E(\bfK^2)}(1+\bfn\cdot\bfs)\vert\psi_0(\bfK^2)\vert^2\label{protch}\\
&&{\rho}_Q^{GAn}(\bfK,\bfn,\bfs)=
{-\vert\psi_0(\bfK^2)\vert^2\over 2E(\bfK^2)}\left[(1+\bfn\cdot\bfs)
+{{\cal R}^2\over 3}\left(-3-2{\bfK^2\over m_s^2}+\bfn\cdot\bfs-{2\over  m_s^2}
\bfs\cdot\bfK\bfn\cdot\bfK\right)\right]\nonumber\\
&&{\rho}_1^{GA}(\bfK,\bfn,\bfs)=
{3\vert \psi_0(\bfK^2)\vert^2\over 2E(\bfK^2)}\left[(1+\bfn\cdot\bfs)
-{{\cal R}^2\over 3}\left(-3-2{\bfK^2\over m_s^2}+\bfn\cdot\bfs-{2\over  m_s^2}
\bfs\cdot\bfK\bfn\cdot\bfK\right)\right].\label{gashape}
\eea 
There are a number of noteworthy features. The non-spherical nature of the
distributions arise only from the vector di-quark part of the wave function.
This does not contribute to the proton charge and spin-dependent charge densities,
leading to a spherical (but polarization dependent) result. The charge distribution 
of the neutron and the matter distribution of the proton and neutron are each
non-spherical with the degree of non-sphericity controlled by the terms
${\cal R}$ and ${\bfK^2\over m_s^2}$.

Note that \eq{protch} 
corresponds to a charge distribution of the proton because the charge
operator is of the explicit form: $1/6+\tau_3/2$ in \cite{ga}. That work
accounts for the effects of the pion cloud \cite{cbm,lfcbm} in an 
approximate manner. In general,
charge would be carried by both the quarks and the pion. In that case the 
quark charge operator would take the form $f_++f_-\tau_3$, and the
vector diquark term would contribute a non-spherical density proportional to 
 $3f_+-f_-.$
 In general, quark lines are dressed by loops
and the dressing can be applied to the quark charge density as shown in  \cite{KB6}.

\begin{figure} [t!]
\unitlength1cm
\begin{picture}(10,8)(4,1)
\includegraphics{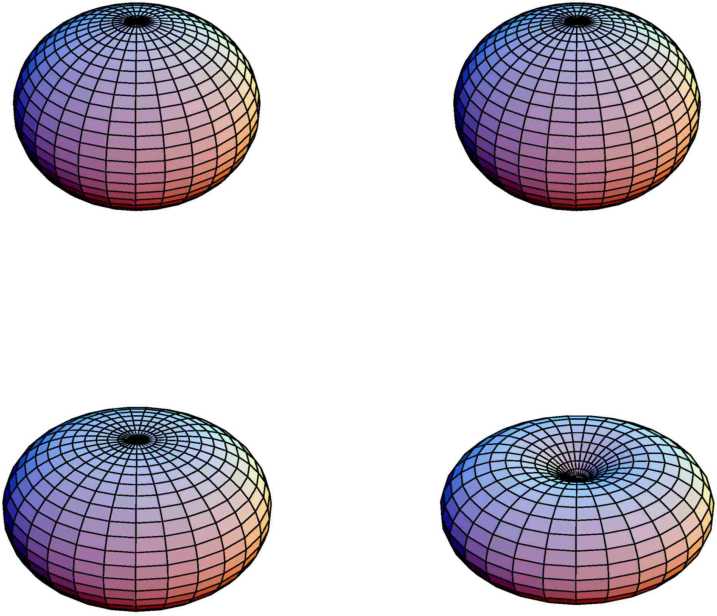}%  angle=0 hscale=75 vscale=75}
\end{picture}
\label{fig:anti}
\caption{Matter distributions. The spin direction $\bfs$ is taken as  vertical,
$\bf{\hat {z}}$, with
$\bfn=-\hat{\bfs}$. Right column model of \cite{Miller:2003sa}  
, left column
model of \cite{ga}.} %First row $K=250 MeV/c$, second row $K$=1 GeV/c.}
\end{figure}
%\vskip10.2cm
\section{Comparing Shapes}
We turn to the numerical evaluation and display of these shapes, focusing on
a comparison of the matter distributions  of the Miller \cite{Miller:2003sa} and
GA models. Specific numerical evaluations of the shapes requires knowing the
value of $m_s$, which is not specified in Ref.~\cite{ga}. Here we note that in
the non-relativistic limit ${\cal R}\approx 1$ and the quantities $\bfK^2/m_s^2$ and 
$\gamma$ play the same roles. In that case, we simply set $\bfK^2/m_s^2=\gamma$.
We also consider a case when the quarks are moving relativistically with
$K=1$ GeV/c. Then $\gamma=0.6$. The relation between $K$ and $\gamma$ is
$K=2m\gamma/(\gamma^2-2 \gamma+1).$ %\to m_s\gamma/(\gamma^2-2 \gamma+1)$. T
It is reasonable to associate the values of $2m$ with the diquark mass
$m_s$. In this case,
the value  
$\gamma=0.6$  corresponds to $K^2/m_s^2=3.75,\;\chi=2.36,\;{\cal R}=0.53.$ 
These numerical values are used to obtain the figures shown below.

The first situation  we consider is the one in which the quark spin is
parallel to the nucleon spin (total angular momentum). 
In the model \cite{Miller:2003sa} the quark probability is larger at the 
top than at the side by a factor of $1+\gamma$. In the model \cite{ga} the corresponding
factor is 1.24 for quarks of momenta 1 GeV/c. See  Fig.~1. %\ref{fig:para}.
In the limit of $K\to\infty$ the ratio would be 2 in both models.
Thus the peanut configuration is hiding within the nucleon of \cite{ga}, albeit with
very small probability.

Next we take the quark spin anti-parallel to the nucleon spin: $\bfn=-\hat{\bfs}$,
see Fig.~2. %\ref{fig:anti}.
In this case the non-sphericity of the two models is very similar.
Both have a toroid shape, with the model \cite{ga} having more pronounced effects.

%\newpage

%\vskip4.0cm
\begin{figure} [t!]
%\begin{Large}
\unitlength1cm
\begin{picture}(10,8)(0,1)
\includegraphics{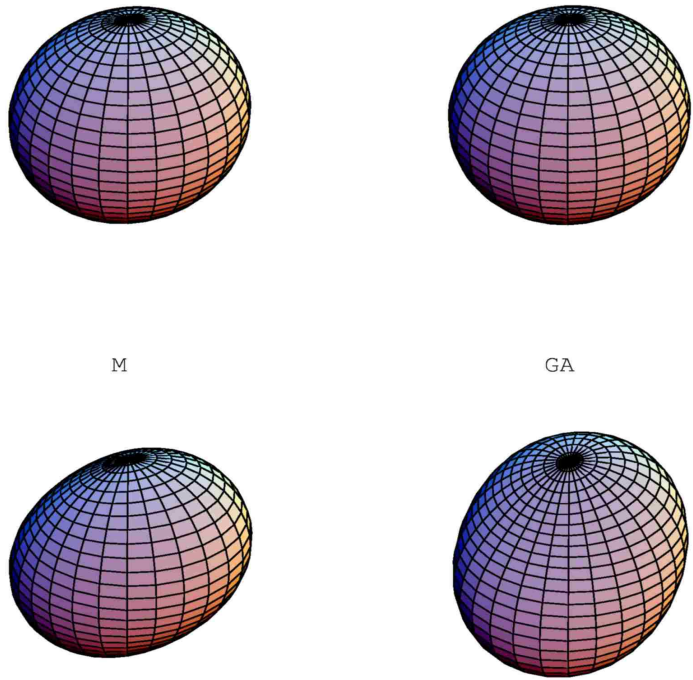}
\end{picture}
%\end{Large}
\label{fig:side}
\caption{Matter distributions. The spin direction $\bfs$ is taken as  vertical,
$\bf{\hat {z}}$, with
$\bfn\cdot\hat{\bfs}=0$. The direction of  $\bfn$ is in the page ($y$ direction).
 Right column model of \cite{Miller:2003sa}, left column
model of \cite{ga}. First row $K=250$  MeV/c, second row $K$=1 GeV/c.}
\end{figure}

The final situation we consider is that in which the quark spin is perpendicalar to
that of the nucleon spin, see Fig.~3. We take $\bfn\cdot\bfs=0$ with
$\bfn$ pointing along the right side of the figure. Once again each model nucleon
has a  significant 
deformation. 

\section{Summary}
The nucleon is far from  round in  each  of the models considered,
and this arises from the relativistic nature of each. The deviation from a spherical
shape is    associated physically with motion of  spin 1/2 
quarks moving relativisitically within the nucleon, and 
mathematically with the  non-vanishing of the term 
$C_{\cal O}(\bfK^2)$ of \eq{genshape}. The general 
nature of $C_{\cal O}(\bfK^2)$ is a subject for future work.

\section*{Acknowledgments}
We thank the USDOE and the CRDF grant 
 GEP2-3329-TB-03  for partial support of this work.

\end{document}